\documentclass[a4paper,fleqn,final]{cas-dc}

\usepackage[numbers]{natbib}
\usepackage{subfigure}
\setcitestyle{numbers,square}

\def\tsc#1{\csdef{#1}{\textsc{\lowercase{#1}}\xspace}}
\tsc{WGM}
\tsc{QE}
\tsc{EP}
\tsc{PMS}
\tsc{BEC}
\tsc{DE}

\graphicspath{{Figure/}}

\begin{document}

\let\WriteBookmarks\relax
\def\floatpagepagefraction{1}
\def\textpagefraction{.001}

\shorttitle{Coevolution under recommendation protocol}

\shortauthors{Hongyu Yue et~al.}


\title {Coevolution of relationship-driven cooperation under recommendation protocol on multiplex networks}

\author[1]{Hongyu Yue}[style=Chinese]

\author[1]{Xiaojin Xiong}[style=chinese]

\author[1]{Minyu Feng}[style=chinese,orcid=0000-0001-6772-3017]
\cormark[1]
\ead{myfeng@swu.edu.cn}

\author[2]{Attila Szolnoki}

\affiliation[1]{organization={College of Artificial Intelligence},
    addressline={Southwest University}, 
    city={Chongqing},
    postcode={400715}, 
    country={PR China}}

\affiliation[2]{organization={Institute of Technical Physics and Materials Science},
    addressline={Centre for Energy Research}, 
    city={Budapest},
    postcode={H-1525},
    country={Hungary}}

\begin{abstract}
While traditional game models often simplify interactions among agents as static, real-world social relationships are inherently dynamic, influenced by both immediate payoffs and alternative information. Motivated by this fact, we introduce a coevolutionary multiplex network model that incorporates the concepts of a relationship threshold and a recommendation mechanism to explore how the strength of relationships among agents interacts with their strategy choices within the framework of weak prisoner’s dilemma games. In the relationship layer, the relationship strength between agents varies based on interaction outcomes. In return, the strategy choice of agents in the game layer is influenced by both payoffs and relationship indices, and agents can interact with distant agents through a recommendation mechanism. Simulation of various network topologies reveals that a higher average degree supports cooperation, although increased randomness in interactions may inhibit its formation. Interestingly, a higher threshold value of interaction quality is detrimental, while the applied recommendation protocol can improve global cooperation. The best results are obtained when the relative weight of payoff is minimal and the individual fitness is dominated by the relationship indices gained from the quality of links to neighbors. As a consequence, the changes in the distribution of relationship indices are closely correlated with overall levels of cooperation.
\end{abstract}


\maketitle

\section{Introduction}
Throughout the development of human civilization, conflict and struggle were pervasive, yet cooperative behavior remained widespread in both natural and social systems \cite{nowak_11,sigmund_10}. The emergence of cooperation attracted considerable scholarly attention due to its profound implications in various fields, such as economics, biology and the social sciences\cite{capraro2024outcome, feng2024information}. As a result, the development of evolutionary game theory provided a solid theoretical foundation and an effective framework for studying this phenomenon. Within this framework, the so-called Prisoner's Dilemma Game \cite{SZOLNOKI2020109447, zhu2020investigating, perc2008social}, Snowdrift Game \cite{li2021pool, hauert2004spatial, xia_cy_ps11}, and Stag Hunt Game \cite{luo2021evolutionary, starnini2011coordination, SILVA2024102290} represent different aspects of the original dilemma and they were studied extensively. As an important step toward more realistic models, several studies were conducted on spatial populations by using complex networks, including square lattice networks \cite{nowak_n92b, szabo_pre05}, small-world networks \cite{weeden2020small, latora2001efficient}, and scale-free networks \cite{santos_prl05, szolnoki_pa08, chen_xj_pla08}. In these networks, each node represented a single player, while each edge represented their interaction. Players engaged in specific types of games with their neighbors and updated their strategies based on their respective payoffs. In general, both the game model and the network topology had a significant impact on the evolution of cooperative behavior \cite{fu_pa07}.

In recent years, coevolutionary game theory emerged as a crucial framework for understanding the evolution of cooperative behavior \cite{perc_bs10}. In these models, the evolution of individual strategies influences other elements of the model which also determines the microscopic dynamics. This can be some player specific feature, like strategy teaching or learning capacity \cite{szolnoki_njp08}, reputation \cite{xia_cy_c20}, morality\cite{capraro2021mathematical}, or the microscopic environment of an agent describing proper interactions with neighbors \cite{chen_xj_pcb18}. Evidently, the latter includes the application of network science \cite{boccaletti_pr06, boccaletti_pr14, guo_h_jrsif20}. 

Complex networks developed rapidly, leading to the emergence of various novel network models in addition to the basic single-layer network model \cite{feng2015evolving, feng2018evolving}, including multilayer networks \cite{wang_z_epl12, kivela2014multilayer,arruda_pr18}, temporal networks \cite{holme2012temporal, li2023evolving, zeng2023temporal}, and higher-order networks \cite{alvarez-rodriguez_nhb21, benson2016higher}. These novel network models laid the foundation for research on spatial evolutionary games. Multilayer networks allow different types of interactions to coexist within the same system, such as various transportation modes in transportation networks \cite{aleta2017multilayer} and multidimensional interactions in social networks \cite{dickison2016multilayer}. This network structure, by introducing interactions at different levels, provided a more realistic dynamic analytical framework that played a crucial role in the study of the evolution of cooperative behavior. For instance, in evolutionary games, cooperative and defection behaviors were influenced not only by direct interactions between individuals but also by potential propagation through multiple levels of indirect interactions \cite{weibull1995evolutionary, xiong2024coevolution}. Based on the nature of network connections and interactions, multilayer networks could be further classified into multiplex networks \cite{halu2013multiplex, gomez2013diffusion}, interdependent networks \cite{gao2012networks}, and interconnected networks \cite{duato2002interconnection}. Additionally, the application of multilayer networks in network epidemic propagation models became increasingly widespread \cite{feng2023impact, 7093190}, enabling researchers to more authentically explore the role of multiple information interactions in social phenomena.

In the real world, networks are not only multilayered but also exhibit coevolutionary characteristics. Agents typically engaged in interaction selection based on social relationships before participating in evolutionary games, rather than simply interacting with all neighbors. Most existing studies assumed that the probability of agents engaging in games was uniform across the network. However, this assumption neglected the potential influence of the strength of relationships between agents on the probability of interaction. Based on this, we construct a coevolutionary multiplex network model, which displays the structure of multiplex networks and introduces a relationship layer to represent interpersonal connections between agents. Before participating in games, agents first establish relationships with other nodes through this layer. The probability of each agent engaging in a game is closely related to the strength of their relationships in the relationship layer. When the relationship strength exceeds a certain threshold, the agent has a stronger willingness to engage in a game with that neighbor. During the coevolutionary process, the strength of relationships between agents is influenced by the outcomes of each game, that is, cooperation strengthens relationships, while defection declines them. As a key novel element of our description, we introduce a recommendation mechanism, whereby each agent could recommend neighbors with whom they have a strong relationship to their optimal neighbors, provided that the relationship strength exceeds a certain threshold. It means that even if some nodes are not directly connected in the relationship layer, they have the probability to engage in indirect interactions in the game layer due to recommendations, thereby fostering a broader cooperative network.

This study is organized as follows. In Section~\ref{sec:model}, we describe our model: the coevolution of the multiplex network and the strategy updating rules based on improved fitness. In Section~\ref{sec:results}, we detail our experimental results on different network structures. Finally, in Section~\ref{sec:conclusion}, we conclude with a summary of our findings and outline prospects.

\section{Model}
\label{sec:model}

In interpersonal interactions, varying degrees of relationship strength between agents influence their decision-making behavior. Furthermore, agents may exhibit recommendation behavior when their relationships are particularly strong. In this section, we introduce a coevolutionary model based on the Weak Prisoner's Dilemma (WPD) \cite{nowak_n92b}, incorporating agent relationships and a recommendation mechanism within a multiplex network framework. To achieve this, we construct a multiplex network consisting of a relationship layer and a game layer. We assume that each agent has varying degrees of relationship strength with others, and when this strength surpasses a certain threshold, agents are more likely to engage in games, accompanied by the emergence of recommendation behavior. Our primary focus is to investigate the interaction between these two network layers and the mechanism of their coevolution.

\subsection{Network Topology and Adaptive Relationship Strength}
In a conventional prisoner's dilemma, each player is presented with two distinct choices: cooperation (C) or defection (D). During every round of interaction, both players simultaneously and independently select their strategies. If both players decide to cooperate, they each receive a reward, labeled as $R$. On the other hand, if both choose defection, they each get a payoff represented as $P$. When one player defects while the other cooperates, the defector secures a higher payoff, $T$, while the cooperator receives a lower payoff, $S$. The condition \(T > R > P > S\) ensures that defection remains more rewarding than cooperation, regardless of the other player's decision. To maintain the core challenge of the social dilemma while simplifying the model, we adopt the WPD parameters, where $R$ is set to 1, $T$ is represented by $b$, and both $S$ and $P$ are set to 0, with $b$ varying between 1 and 2.

Based on WPD, we construct a multiplex network to capture the process of coevolution, consisting of a relationship layer and a game layer.
In the relationship layer, each node represents an agent, and there exist edges between them. The weight of an edge indicates the strength of their relationship by a $W$ value ranging from [0, 1]. 
In the game layer, the agents correspond to the same nodes as in the relationship layer. At each time step, agents play the WPD game with their neighbors with different probabilities. If the relationship strength between agents \(x\) and \(y\) in the relationship layer exceeds the threshold \(\gamma\), the probability of them participating in the game is \(p\), otherwise, it is \((1 - p)\).

Importantly, the agents' decision-making behavior in the game layer also affects the weight of edges in the relationship layer. We define the change in relationship rate as \(\delta\). If both agents in the game choose to cooperate, the edge weight in the relationship layer increases by \(\delta\). On the contrary, if both agents choose to betray, the corresponding edge weight will also decrease by \(\delta\). If only one agent chooses to cooperate, the relationship strength remains unchanged. To describe how the decision-making behavior of the agents in the game layer affects the strength of the relationship in the relationship layer, we define the following formula

\begin{equation}
W(i,j) = 
\begin{cases} 
[W(i,j) + \delta]^1_0\,, & \text{if } S_i = S_j = C \\
[W(i,j) - \delta]^1_0\,, & \text{if } S_i = S_j = D  \\
W(i,j)\,, & \text{otherwise}
\end{cases},
\end{equation}
where \( W(i,j) \) represents the relationship strength between agents \(i\) and \(j\), and \( S_i \) and \( S_j \) denote their respective decision-making behaviors. A constraint \([ \cdot ]_{0}^{1}\) is used to ensure that the relationship strength remains within the range \([0,1]\), that is

\begin{equation}
[a]_{0}^{1} =
\begin{cases}
0, & \text{if }  a < 0 \\
a, & \text{if }  0 \le a \le 1 \\
1, & \text{if }  a > 1
\end{cases}.
\end{equation}

\subsection{Recommendation Mechanism and Strategy Evolution}
In reality, we do not only interact with our stable partners but sometimes with others. These temporal links, however, are not purely random but based on a recommendation of our partners. Accordingly, the recommendation mechanism introduces a new mode of interaction for agents in the game layer, where relationships not only influence direct interactions but also extend to indirect interactions through third-party recommendations. In the relationship layer, each agent identifies a preferred or optimal neighbor, defined as the neighbor with whom they share the strongest relationship. This close connection creates opportunities for indirect participation in the game layer.

When the relationship strength between agent $x$ and agent \(y\) in the relationship layer exceeds the threshold \( \gamma \), they engage in a game with the probability \(p\). However, the recommendation mechanism allows agent \(x\) to recommend their optimal neighbor \(z\) to interact with agent \(y\), even in the absence of a direct connection between \(z\) and \(y\). This recommendation also occurs with the probability \(p\).

The mechanism extends the possibilities for interactions in the game layer, making indirect cooperation more likely, thereby influencing the overall system dynamics. Although the recommendation is probabilistic, it adds complexity to the model by affecting not only the immediate outcomes of the game but also the evolution of relationship strength in subsequent iterations. Through recommendations, agents can indirectly shape the social structure and cooperation patterns within the network, potentially leading to more dynamic system behavior and higher levels of global cooperation.

It is a key element of our coevolutionary approach that the relationship layer affects the evolution of strategies. Traditionally, the payoff is the main determinant of fitness. However, in this model, the strength of interpersonal relationships in the relationship layer is also an important factor. To quantify this, we introduce a relationship index as part of the fitness weight. We denote the relationship index of node \(i\) as

\begin{equation}
A_i = \sum_{j \in V} W(i,j)\,,
\end{equation}
where \( V \) denotes the set of neighbors of node \( i \) in the relationship layer, and \( W(i,j) \) represents the relationship strength between nodes \( i \) and \( j \). The relationship index is determined by the decision-making behaviors between a node and its neighbors. Therefore, compared to reputation, the relationship index can be viewed as a form of ``dynamic fitness'' or ``interaction weight''. It not only reflects the relationship strength of a node in the relationship layer but also captures the strategic influence arising from interactions with neighbors. The relationship index can reveal a node's relative advantage or disadvantage in the decision-making process, rather than merely reflecting the static social status.

As a plausible assumption, nodes with stronger relationships with other nodes in the relationship layer generally have higher fitness. Here, we assume that both payoff and relationship index are directly proportional to fitness. Therefore, the fitness of node \( i \) can be expressed as
\begin{equation}
f_i = m \Pi_i + A_i\,,
\end{equation}
where \( \Pi_i \) represents the payoff of the node \( i \) at the current time step, and \( m \) is a damping factor used to balance the relative impact of payoff and relationship index, ranging between 0 and 1.

After each round of the game, players can choose neighbors in the game layer to imitate their strategies. The probability of selecting a neighbor is given by
\begin{equation}
P(i \to j) = \frac{W(i,j)}{A_i}\,.
\end{equation}

Next, considering that interpersonal relationships also affect agent strategy evolution, we apply an imitation process based on a pairwise comparison of fitness \( f_i \) values in the Fermi-function. Accordingly, the probability that node \( i \) adopts the strategy of node \( j \) is given by
\begin{equation}
\Gamma(S_i \leftarrow S_j) = \frac{1}{1 + \exp\left(\frac{f_i - f_j}{\kappa}\right)}\,,
\end{equation}
where \( \kappa \) measures the stochasticity or noise level in the strategy selection process. As \( \kappa \to \infty \), the strategy update becomes random, indicating that individuals are less sensitive to differences in fitness. Conversely, as \( \kappa \to 0 \), individuals always choose the option with higher fitness, making the selection more rational.

\begin{figure*}[htbp]
    \centering
    \includegraphics[bb=-0 -0 611 384, width=0.8\textwidth]{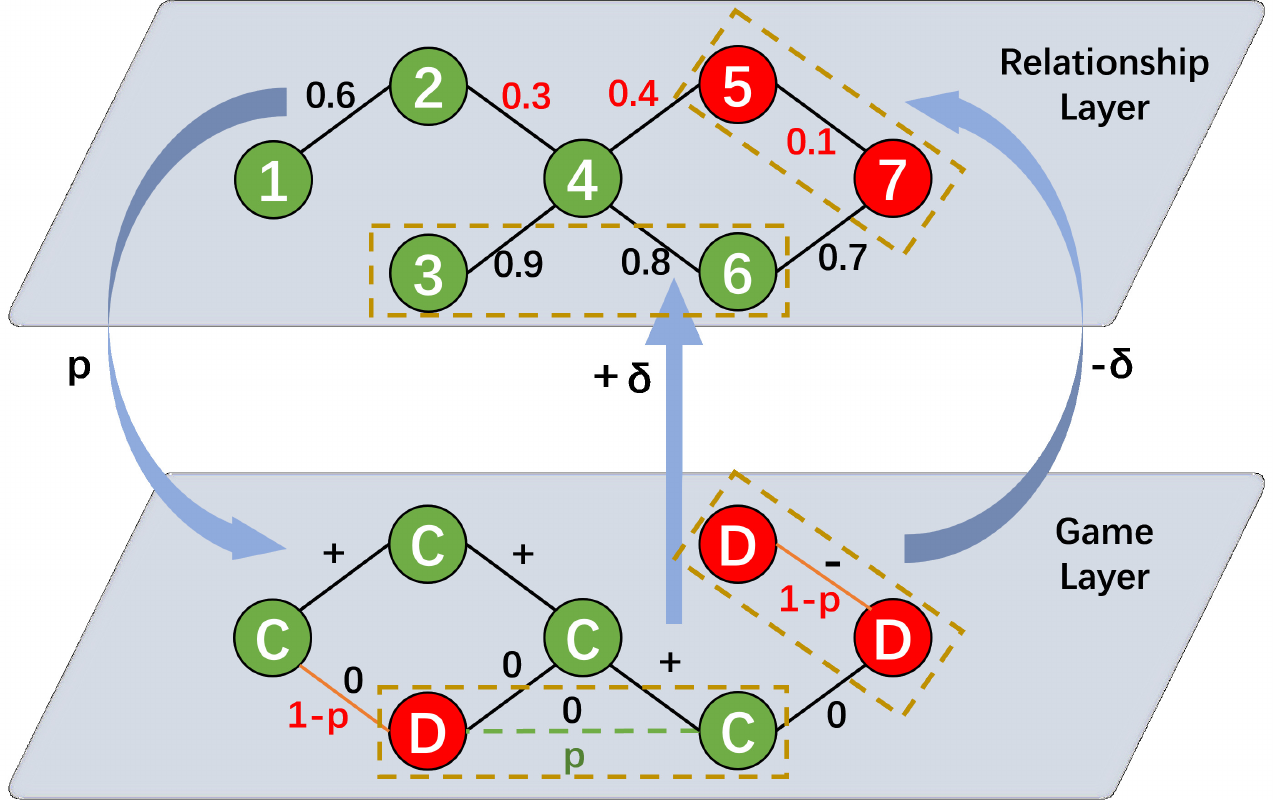}
    \caption{\footnotesize \textbf{Coevolutionary game on multiplex networks.} The model consists of a multiplex network, namely the relationship layer and the game layer. The nodes in both layers correspond to the same agents. In the relationship layer, the edge weights represent the strength of the associations between agents, while in the game layer, the edges indicate the interactions or games played between agents. According to the coevolutionary rules, the strength of associations in the relationship layer influences the interactions in the game layer, and the outcomes of these interactions subsequently affect the association strength in the relationship layer. Further details can be found in the main text.}
    \label{fig1}
\end{figure*}

The details of our model are summarized in Fig.~\ref{fig1}, where agents have different degrees of association with each other, represented by diverse edge weights. The nodes in both the game layer and the relationship layer correspond to the same agents. In the relationship layer, each agent possesses a relationship index that reflects the strength of its connections within the layer and captures the strategic influence arising from interactions with neighbors. In the game layer, agents have only two possible strategies, cooperation or defection. It is noteworthy that an agent's neighbors in the two layers may be different, depending on the likelihood of interaction in the game, which is determined by the probability \(p\) in the relationship layer. When the relationship strength between an agent and its neighbor in the relationship layer exceeds a threshold \( \gamma \), or when there is a recommendation behavior from an optimal neighbor, the agents will engage in game interaction with a probability \(p\); otherwise, the game will proceed with the probability of \((1-p)\). For instance, agents 5 and 7 have a relationship strength below the threshold \( \gamma \) in the relationship layer, so they establish a link in the game layer with a probability of \((1-p)\) (orange solid line). Similarly, although agents 3 and 6 do not have a direct connection, due to agent 4’s recommendation, they will interact in the game layer with a probability of \(p\) (green dashed line). As described in Section 2.1, the strategies of neighbors also influence the relationship strength between agents. In the game layer, when both agents choose to cooperate (represented by green vertices), the edge between them is marked with a '+', indicating that the weight of their corresponding edge in the relationship layer will increase by \( \delta \). Conversely, an edge marked with '-' implies that the relationship strength will decrease by \( \delta \). Edges marked with '0' indicate that their relationship remains unaffected. As explained above, the behavior of agents in the game layer evolves according to fitness-based strategy update rules.

\subsection{Applied graphs forming multiplex networks}
In the following, we analyze the coevolution of the cooperation level and the relationship index under different game parameters. To investigate the potential influence of the relationship layer on system dynamics, we employ various network topologies. Specifically, we apply a Honeycomb Lattice (HL) with a degree of \(k = 3\), a Square Lattice (SL) with \(k = 4\), and a Hexagonal Lattice (XL) with \(k = 6\) with periodic boundary conditions. Additionally, we incorporate the Watts–Strogatz Small-World network (WS) with a degree \(k = 10\) and a rewiring probability \(p = 0.5\). For consistent comparison, all networks are composed of \(N = 2500\) nodes. Initially, each agent is randomly designated as either a cooperator or a defector with equal likelihood. To achieve the stationary state of the coevolutionary process, we execute a total of 10,000 Monte Carlo (MC) steps for each parameter, where the last 1,000 steps are used to measure system-specific quantities, such as the fraction of cooperators or the mean relationship index. All simulation procedures are implemented using Python 3.10.
\section{Analysis of the simulation results}
\label{sec:results}

\subsection{Heatmap of Cooperation Density}
\begin{figure*}[htbp]
    \centering
    
    \subfigure[HL]
    {
        \includegraphics[bb=-0 -0 576 432, scale=0.3]{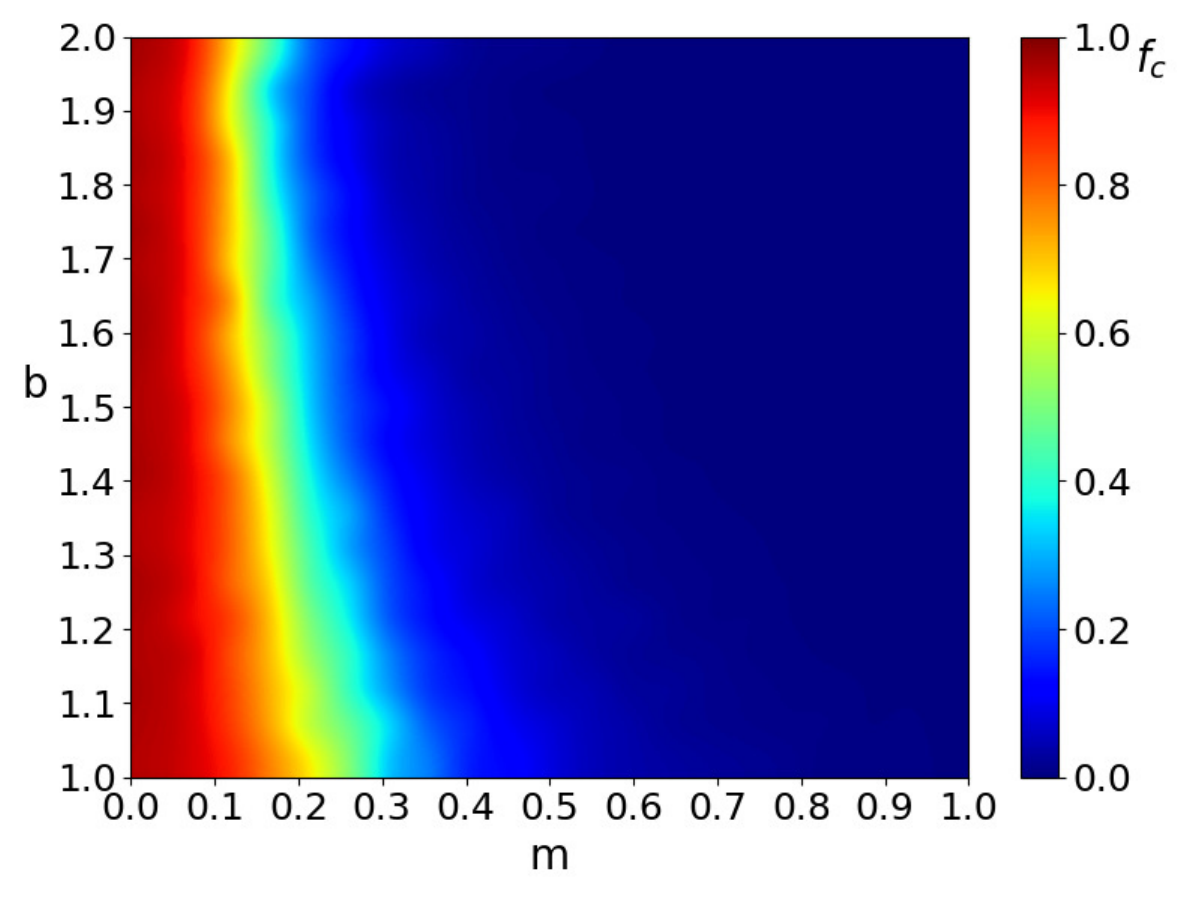}
        \label{fig2:HL}
    }
    \hspace{3mm}
    \subfigure[SL]
    {
        \includegraphics[bb=-0 -0 576 432, scale=0.3]{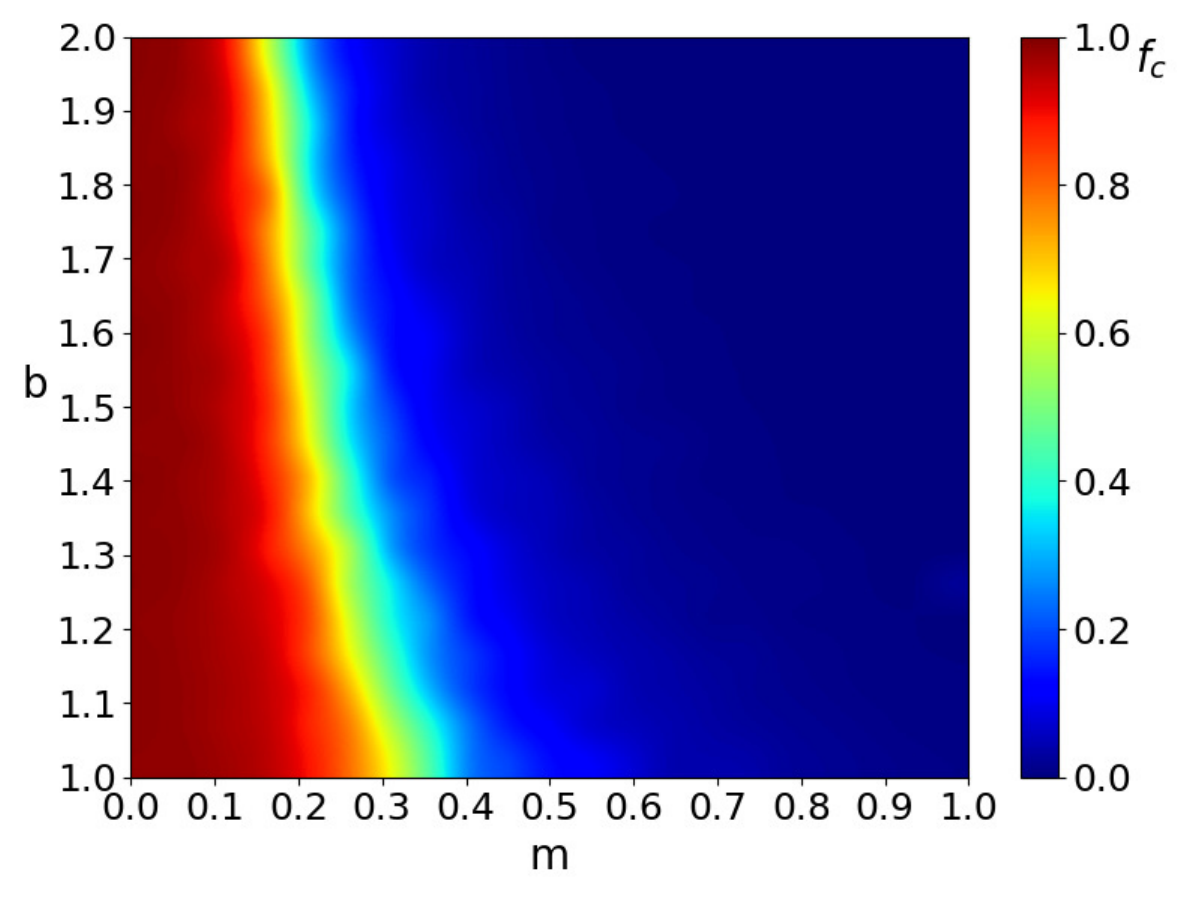}
        \label{fig2:SL}
    }
    \hspace{3mm}
    \subfigure[XL]
    {
        \includegraphics[bb=-0 -0 576 432, scale=0.3]{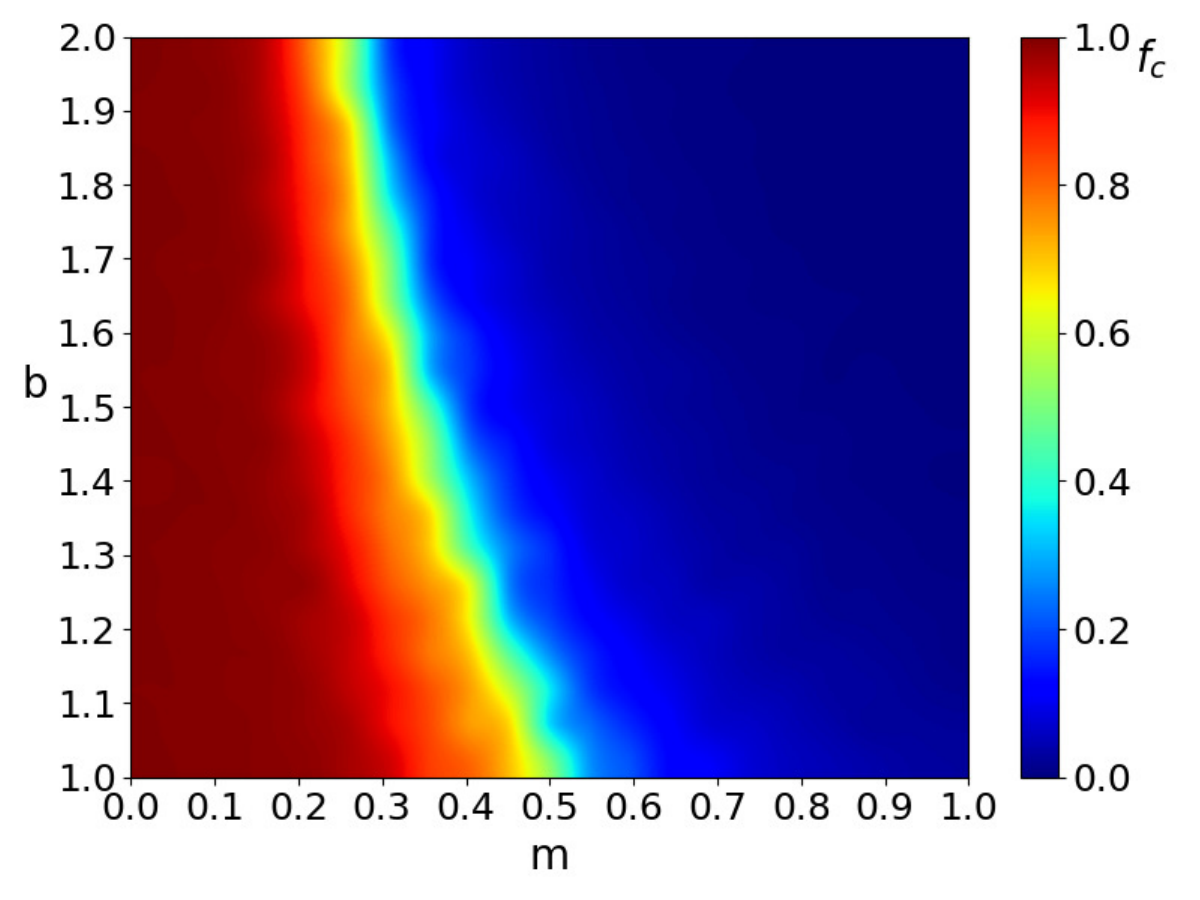}
        \label{fig2:XL}
    }
    \hspace{3mm}
    \subfigure[WS]
    {
        \includegraphics[bb=-0 -0 576 432, scale=0.3]{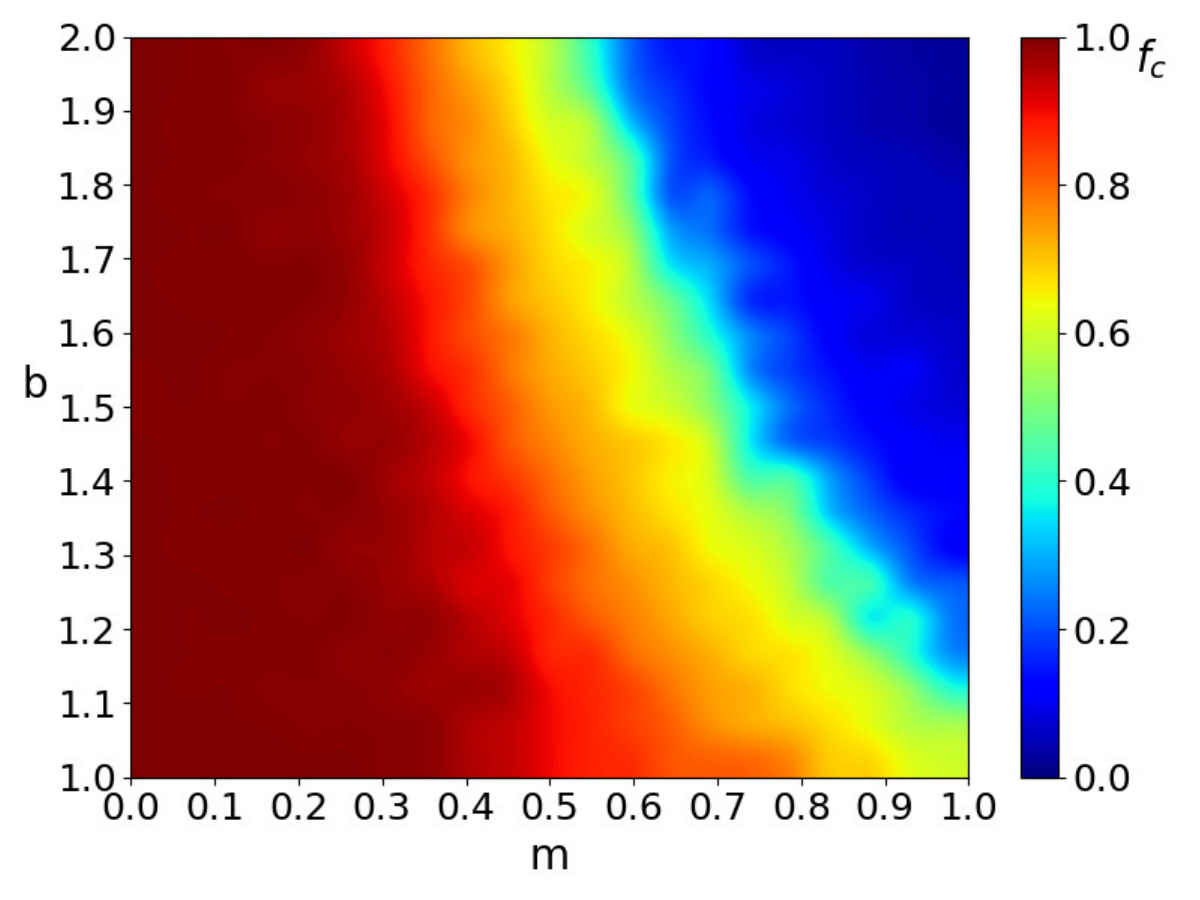}
        \label{fig2:WS}
    }
    
    \caption{\footnotesize \textbf{Cooperation density distribution on $b-m$ parameter plane by using different network types.} The variations in cooperation density $f_c$ concerning parameters $b$ and $m$ are depicted for four network structures: (a) HL, (b) SL, (c) XL, and (d) WS small-world networks, under fixed values of parameter $p = 0.9$ and $\gamma = 0.5$.}
    \label{fig2}
\end{figure*}

By adjusting the temptation parameter \(b\) and the relative weight factor \(m\) in the extended fitness function, we explore the variation in cooperation levels. To gain a more comprehensive understanding of the universality of system behavior, we present results obtained under different network structures specified previously. The typical results are summarized in Fig.~\ref{fig2}, where \(p = 0.9\) and \(\gamma=0.5\) are set in all cases, indicating that agents primarily interact with their neighbors while still having opportunities to engage with other agents with whom they do not have permanent relationships. In all four types of networks, the cooperation density shows a decreasing trend as the parameters \(b\) or \(m\) increase. Within the parameter range examined, we identify areas of pure cooperation and pure defection. Generally, smaller values of \(b\) and \(m\) are associated with higher cooperation rates, while larger values promote an increase in defection behavior. Moreover, the transition from cooperation to defection exhibits a left-skewed trend across all network types.

Within the parameter range examined, we not only identified regions of complete cooperation and total defection but also observed the significant impact of network topology on the evolution of cooperation. For the HL, SL, and XL networks, when $m$ reaches 0.2, 0.3, and 0.5 respectively, the density of cooperators decreases sharply, forming distinct transition lines on the parameter plane, indicating critical points where cooperative behavior undergoes sudden changes. In contrast, the WS network offers more opportunities for the coexistence of cooperation and defection, effectively suppressing the occurrence of pure defection within the population. Across the parameter range studied, lattice networks (HL, SL, XL) are dominated by defection, with cooperation density decreasing rapidly; however, in the WS network, cooperation persists over a broader range of parameters, with pure defection occurring only when both $b$ and $m$ reach high values. It suggests that a higher average degree in the network can not only effectively suppress defection, but also promote the development of cooperation.

\subsection{The Cooperation Level Under Different Parameters}
\begin{figure*}[htbp]
    \centering
    
    \subfigure[]
    {
        \includegraphics[bb= -0 -0 144 93, width=0.3\textwidth]{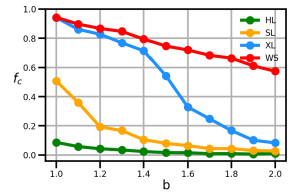}
        \label{fig3a}
    }
    \hspace{3mm}
    \subfigure[]
    {
        \includegraphics[bb= -0 -0 144 93, width=0.3\textwidth]{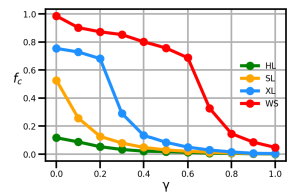}
        \label{fig3b}
    }
    \hspace{3mm}
    \subfigure[]
    {
        \includegraphics[bb= -0 -0 144 93, width=0.3\textwidth]{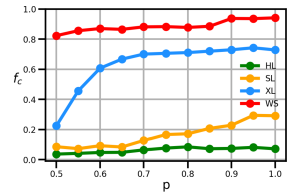}
        \label{fig3c}
    }
    \caption{\footnotesize \textbf{Dependence of cooperation level on different parameters within four networks.} (a) We investigate the dependence of \(f_c\) on \(b\) under fixed conditions where \(m = 0.5\), \(p = 0.9\), and \(\gamma = 0.2\). (b) To investigate the effect of the threshold \(\boldsymbol{\gamma}\), we set \(m = 0.5\), \(p = 0.9\), and \(b = 1.5\) to observe the variation of \(f_c\). (c) The parameters \(m = 0.50\), \(b = 1.5\), and \(\gamma = 0.1\) are employed across four different types of network.}
    \label{fig3}
\end{figure*}

We analyze the variation of cooperation density \(f_c\) with different parameters, as illustrated in Fig.~\ref{fig3}. In particular, Fig.~\ref{fig3a} shows how \(f_c\) varies with respect to the parameter \(b\). The results further support previous observations, indicating that a higher degree within the network enhances cooperation. Within the range \(1 \leq b \leq 2\), the cooperation levels for lattice networks (HL, SL, XL) drop towards zero as \(b\) increases. However, for WS networks, the cooperation density \(f_c\) declines gradually and stabilizes without reaching zero. Notably, both SL and XL exhibit rapid declines at different ranges: SL shows this at smaller values of \(b\) \((1.1 < b < 1.5)\), while XL experiences a sharp drop at larger \(b\) values \((1.7 < b < 2.0)\). As expected, a greater temptation hinders cooperation development, but the rate at which \(f_c\) decays varies across different topologies. In general, graphs with higher average degrees show a stronger resistance to such temptation.

We then observe the impact of the threshold \( \gamma \) on cooperation levels, as shown in Fig.~\ref{fig3b}. It is evident that as \( \gamma \) increases, the cooperation density \( f_c \) approaches zero for all four networks. This indicates that the more stringent the interaction conditions between agents, the fewer individuals are willing to engage in cooperative interactions. Notably, XL and WS exhibit a relatively slow decline in the initial stages as \( \gamma \) increases, but after reaching a certain threshold, they experience a rapid drop. It indicates that even under harsh interaction conditions, increasing the degree of the network can effectively enhance resistance to defection.

In our previous simulations, we establish connections between the relationship layer and the game layer with a fixed probability of interaction, \( p = 0.90 \). To investigate the potential impact of \( p \), which determines the strictness of agents' interactions with neighbors defined in the relationship layer, we conduct simulations by systematically varying its value while keeping \( b = 1.50 \), \( m = 0.50 \), and \( \gamma = 0.1 \) constant. As shown in Fig.~\ref{fig3c}, the variation of \( p \) leads to very different system responses under a specific topology. The results indicate that in WS networks, the cooperative behavior is relatively insensitive to changes in \( p \). Conversely, in HL networks, the level of cooperation remains close to zero regardless of the value of \( p \). In contrast, the level of cooperation in square and hexagonal networks increases as \( p \) increases. Notably, the square network shows a relatively rapid increase in cooperation density when \( p \) is between 0.8 and 0.95. However, the hexagonal network exhibits a more rapid growth at lower values of \( p \), specifically between 0.5 and 0.6. Comparatively, the hexagonal network shows the most significant improvement in cooperation across the entire range of \( p \) from 0.5 to 1.

\subsection{Coevolution of Relationship Index}

\begin{figure*}[htbp]
    \centering
    \includegraphics[bb=-0 -0 598 338, width=\textwidth]{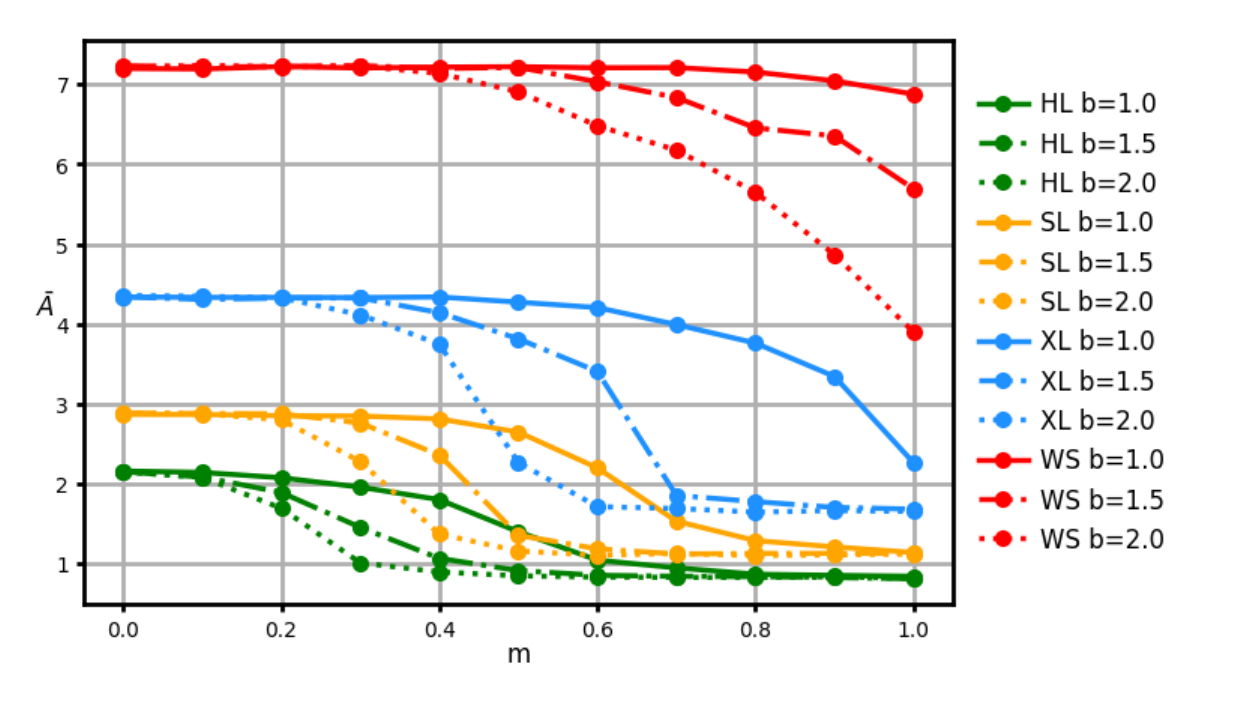}
    \caption{\footnotesize \textbf{The mean of the relationship index \(\bar{A}\) plotted at different \(m\) and \(b\) in four networks.} Different colors represent different network types, as shown in the legend. The solid line, dotted line, and dashed line correspond to the values of \(b = 1.0\), \(1.5\), and \(2.0\), respectively. \( \gamma = 0.1 \) is employed for all four network types.}
    \label{fig4}
\end{figure*}

To gain a deeper understanding of the relationship between cooperation dynamics and network structure, we analyze the average relationship index of all agents in each network as it varies with \(m\) under different \(b\) values, as well as the centrality distribution of relationship indices for all agents under various parameter combinations. The analysis not only evaluates the effects of different conditions on agent behavior but also reveals how to configure the parameter environment to promote the evolution of cooperation, thereby providing a basis for optimizing strategies and enhancing system stability.

\begin{figure}[htbp]
    \centering
    \subfigure[HL]
    {
        \includegraphics[bb=-0 -0 720 432, scale=0.3]{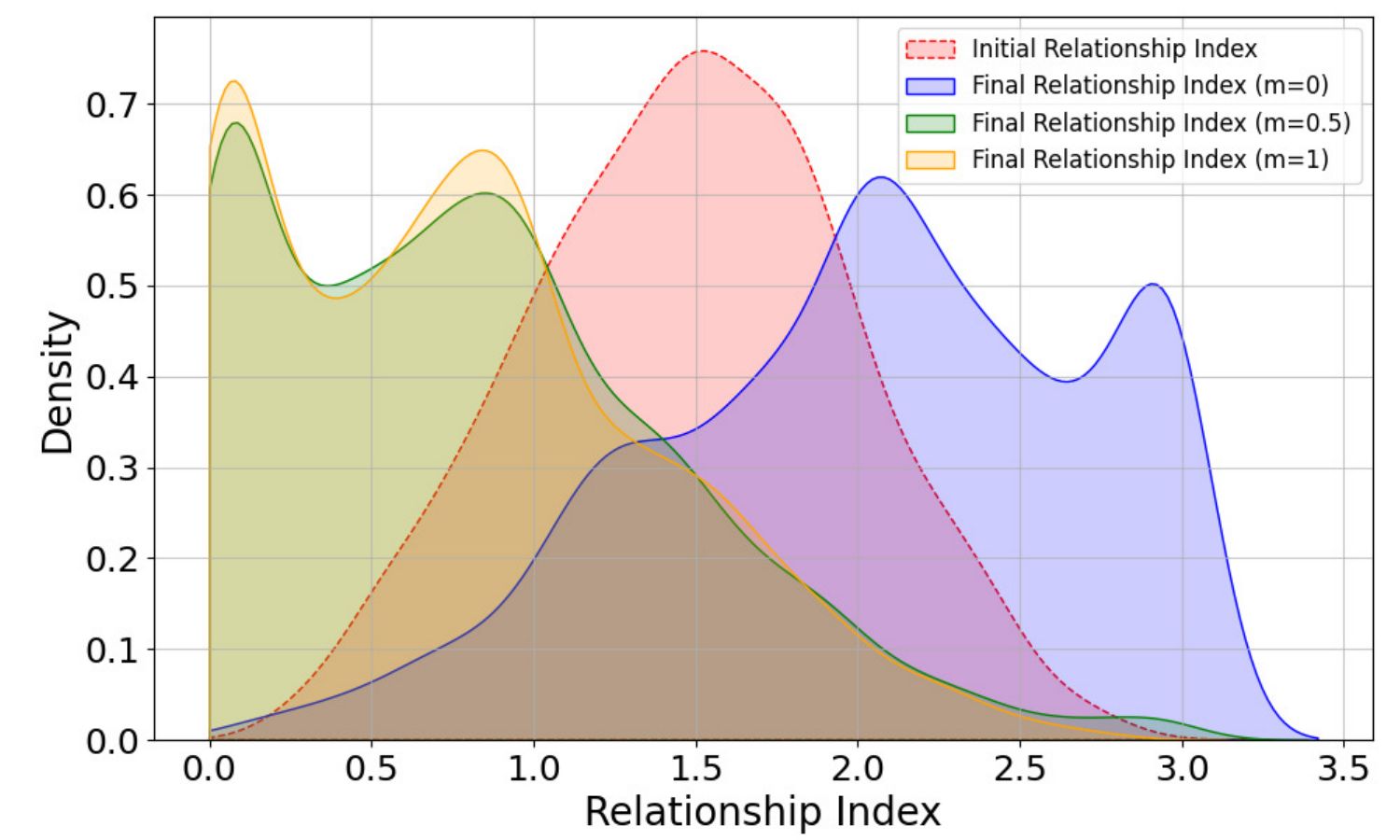}
        \label{fig5:HL}
    }
    \hspace{3mm}
    \subfigure[SL]
    {
        \includegraphics[bb=-0 -0 720 432, scale=0.3]{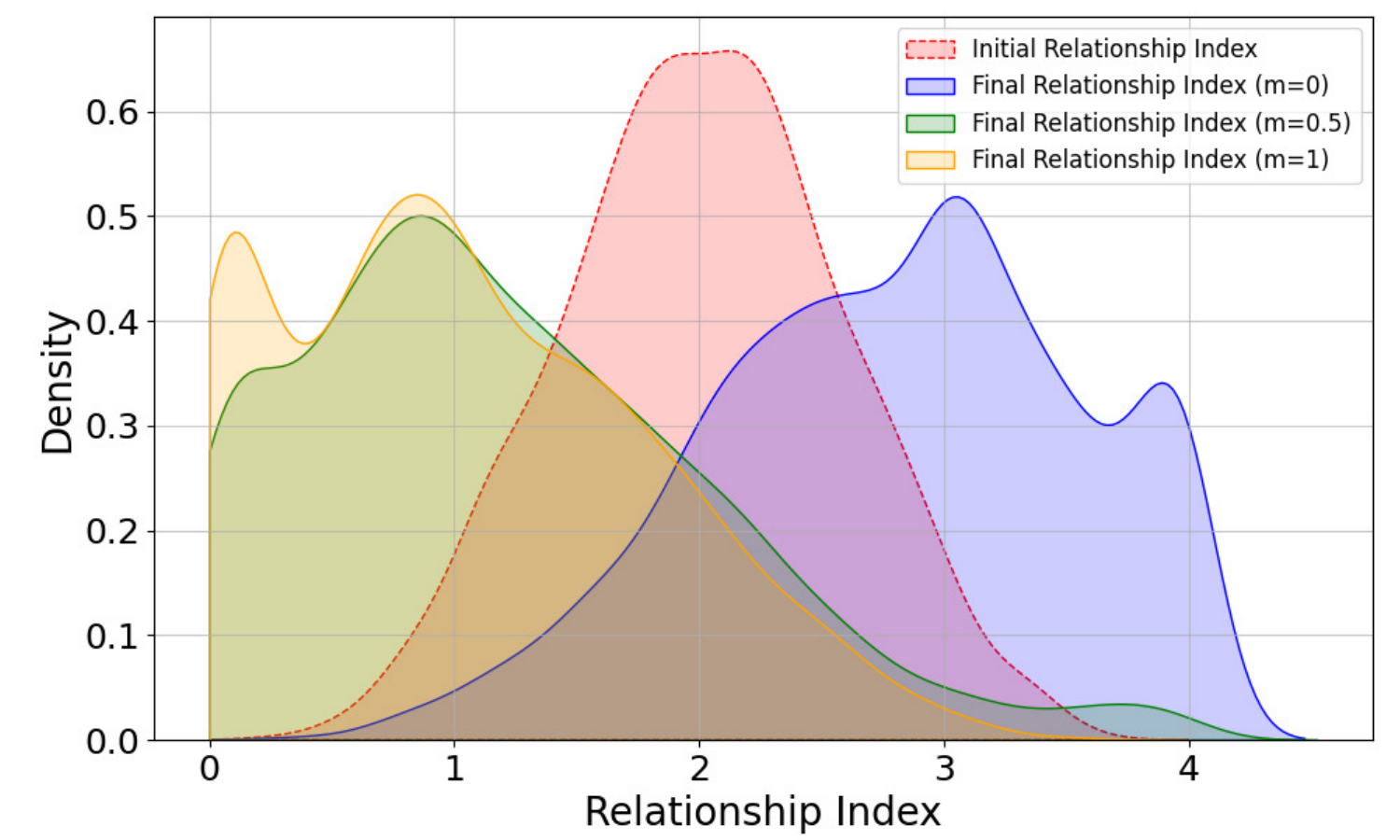}
        \label{fig5:SL}
    }
    \hspace{3mm}
    \subfigure[XL]
    {
        \includegraphics[bb=-0 -0 720 432, scale=0.3]{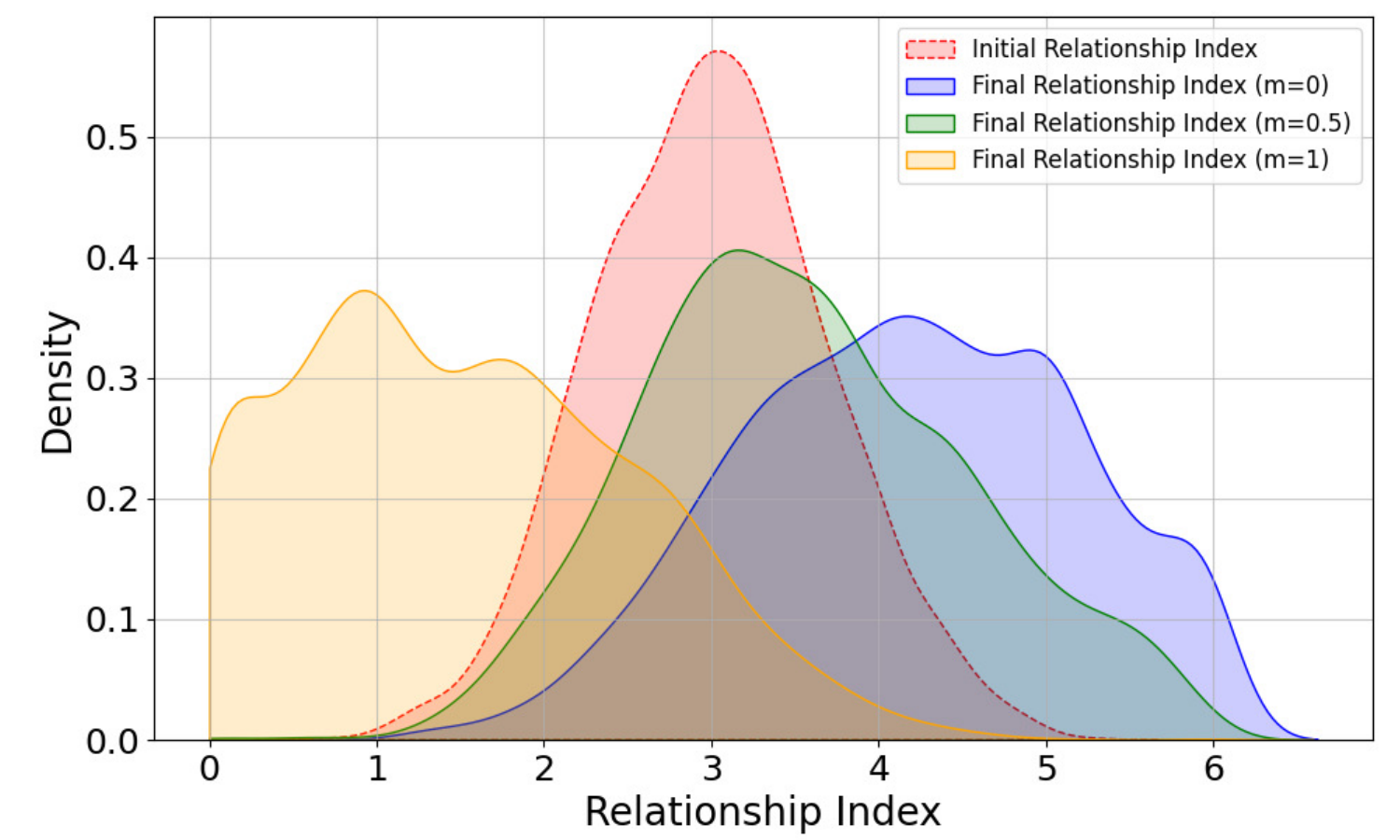}
        \label{fig5:XL}
    }
    \hspace{3mm}
    \subfigure[WS]
    {
        \includegraphics[bb=-0 -0 720 432, scale=0.3]{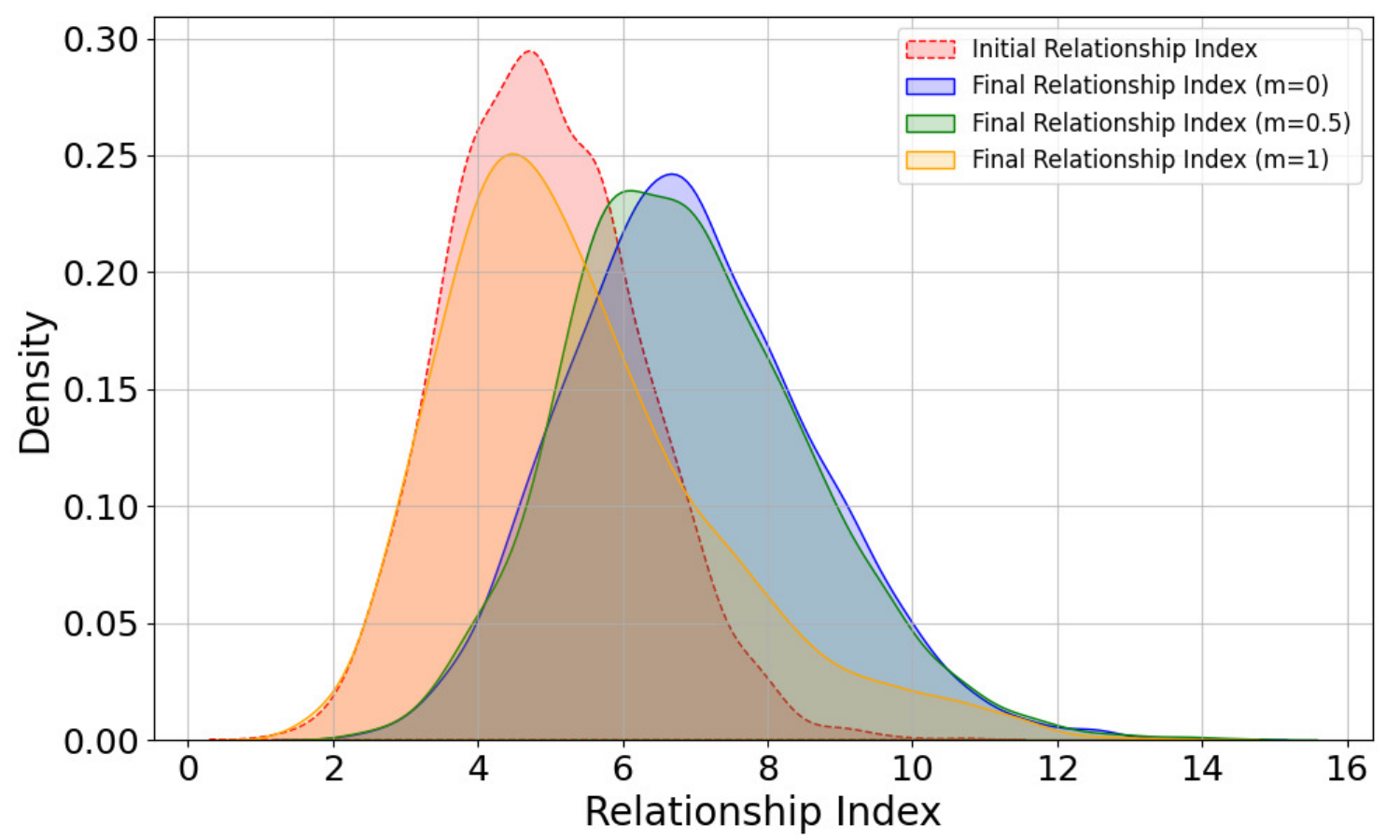}
        \label{fig5:WS}
    }
    \caption{\footnotesize \textbf{Probability density distributions of initial and stationary state relationship indices under different \(m\) values are presented for the four types of networks.} In all cases, \(p = 0.9\), \(b = 1.5\), and \(\gamma = 0.1\) are used. The horizontal axis displays the actual values of the relationship index, while the vertical axis shows their probability density. In all density distribution graphs, the initial normal distribution is represented by a red dashed line, while the steady-state distributions for different \(m\) values are indicated by blue, green, and orange lines, as shown in the legend.}
    \label{fig5}
\end{figure}

As illustrated in Fig.~\ref{fig4}, we examine the relationship between \( m \) and the average relationship index under different temptation values \( b \) (1, 1.5, and 2). Regardless of the temptation intensity, the average relationship index exhibits a marked decrease as \( m \) increases, with this downward trend being particularly pronounced at higher temptation levels. In lattice structures such as HL, SL, and XL, the average relationship index initially declines with increasing \( m \), but eventually stabilizes when \( m \) reaches higher values. Notably, in HL and SL networks, the influence of temptation becomes negligible when \( m \) is sufficiently large, leading the average relationship index to converge to a common stable value. On the other hand, the WS network demonstrates a persistent decrease in the average relationship index, maintaining the highest levels across all networks, with a maximum value approaching 7.2, whereas the upper limit for the HL network is approximately 2.2. In general, the average relationship index increases as the average degree of the network grows, with the WS network consistently showing the highest values.

In addition to examining the average value of \( A \), we further measure the distribution of relationship indices across all networks, as illustrated in the various panels of Fig.~\ref{fig5}. To appropriately evaluate the final steady-state distributions, we also plot the initial normal distribution curves during the analysis. To ensure a fair comparison, we select three representative values of the coupling weight of the fitness function, namely \( m = 0 \), \( m = 0.5 \), and \( m = 1.0 \), while keeping the parameters \( b = 1.5 \), \( p = 0.9 \), and \( \gamma = 0.1 \). The distribution of relationship indices shown in the figure reveals the impact of different network structures on cooperative behavior. Initially, the relationship indices in all networks exhibit a shape close to a normal distribution, which is due to the uniform initial distribution of relationship strength \( W \), reflecting the fact that individuals' social interactions have no significant differences at the start. However, as the game progresses and the evolution of cooperation density \( f_c \) unfolds, we observe a noticeable trend of differentiation. In lattice networks, as the level of cooperation increases, the distribution of relationship indices shifts to the right. It indicates that in these networks, cooperative behavior fosters higher relationship indices, particularly when cooperation density approaches 1, suggesting that the connections between individuals become stronger. Notably, the distributions of relationship indices in the HL and SL networks exhibit two distinct peaks, suggesting the emergence of two different cooperative groups: one maintaining a high level of cooperation and the other gradually reducing cooperation. This bimodal phenomenon reflects the separation between cooperative and defecting groups. In contrast, the distribution of relationship indices in the WS network remains close to a normal distribution throughout, indicating that even as cooperation levels rise, the strength of relationships between individuals does not significantly differentiate. It is attributed to the high stochasticity of connections between individuals in the WS network, leading to less pronounced feedback effects between cooperation and relationship strength compared to the HL or SL networks. Overall, the dynamic process of cooperation and its impact on relationship indices vary significantly across different network structures.

\section{Conclusions and Future Research}
\label{sec:conclusion}
Our study explores agent behavior based on the weak prisoner's dilemma game and the dynamic changes of their relationship networks through a coevolutionary multiplex network model. In this model, agents evolve simultaneously in both the relationship layer and the game layer, with the strength of their relationships and strategy choices mutually influencing each other. Unlike existing studies, our model not only focuses on the interplay between relationship strength and strategy choices among agents but also introduces the concepts of recommendation mechanisms and thresholds to further reveal their impacts on agent strategies.

Our multiplex network consists of a relationship layer and a game layer. In the relationship layer, we declare that the weights of edges between agents represent the strength of relationships, which evolve according to historical interactions and strategy choices, and its value ranges from [0, 1]. In the game layer, agents participate in weak prisoner's dilemma games with neighbors with different probabilities according to their relationships. We introduce a relationship index to quantify the relationships and influence the strategy evolution of the agents. This coupling mechanism realistically simulates the mutual influence between strategy selection and relation dynamics. It is worth noting that we emphasize the role of the recommendation mechanism in our model, where the recommendation is determined by the strength of the relationship, thereby driving the strategy evolution. To comprehensively analyze the cooperative dynamics of the system, this paper tests behavioral evolution under different network topologies, covering honeycomb, square, hexagonal lattices, and Watts-Strogatz small-world networks.

From the previous experimental results, we can get several obvious results. As the temptation parameter \(b\) increases, the global cooperation level decreases significantly, but there are significant differences in the decay patterns of different topological structures. In the HL, SL, and WS networks, the level of cooperation shows a gradual decline, while the XL network exhibits a more pronounced change. In particular, the linear coupling of payoff and relationship index shows a significant negative effect, especially in small-degree networks, which reduces the level of cooperation. This suggests that a simple coupling of the fitness function may not contribute to maintaining cooperation; instead, it may hinder agents' strategic adjustments due to excessive reliance on historical relationships. Our study also reveals that the level of cooperation is closely related to the distribution of the relationship index. In HL and SL networks, cooperation-dominant networks display a right-skewed distribution, while defector-dominant networks exhibit a left-skewed distribution. However, in WS networks, the decline of cooperation is not as pronounced due to the larger average degree, and a trend of transferring to higher relationship values is observed.

 We have studied the phenomenon of recommendation interactions among agents in real life based on relationships in multiplex networks. However, there is still room for improvement in this study. First, the linear coupling in the fitness function is just the simplest approach. Future research can explore more complex coupling methods to model the impact of interpersonal relationships and benefits on fitness more faithfully. Second, this article only considers the close friend recommendation mechanism in the relationship layer. Agents may recommend all neighbors who are related in the relationship layer, even with different probabilities. Finally, in the future, by introducing random variables, we can further study the evolution pattern of cooperation under the conditions of relationship stability and random risk, to have a more comprehensive understanding of the cooperative behavior of agents in real networks.

\section*{Acknowledgments}
H. Yue, X. Xiong and M. Feng are supported by grant No. 62206230 funded by the National Natural Science Foundation of China (NSFC), and grant No. CSTB2023NSCQ-MSX0064 funded by the Natural Science Foundation of Chongqing. A. Szolnoki is supported by the National Research, Development and Innovation Office (NKFIH) under Grant No. K142948.

\bibliographystyle{elsarticle-num-names}
\bibliography{refs}

\end{document}